\documentclass[aps,pre,epsfig,floats,twocolumn,superscriptaddress,amssymb,amsmath,floatfix,showpacs]{revtex4}
\usepackage{graphicx}
\usepackage{bm}  
\usepackage{amssymb}
\usepackage{color}
\usepackage{amscd}

\begin{document}
\title{Casimir force between two substrates within three different layers using the scattering approach}

\author{A. Moradian}
\affiliation{Department of Science, Campus of Bijar, University of Kurdistan, Bijar, Iran}

\author{M. R. Setare}
\email{rezakord@ipm.ir}
\affiliation{Department of Science, University of Kurdistan, Sanandaj, Iran}

\author{S. A. Seyedzahedi}
\affiliation{Department of Science, University of Kurdistan, Sanandaj, Iran}

\date{\today}

\begin{abstract}
We compute the Casimir force for a system composed of two layers as substrates within three different homogenous layers. We use the scattering approach along with the Matsubara formalism in order to calculate the Casimir force at finite temperature.
 We impose the appropriate boundary condition on the tangential components of the electric and magnetic fields to construct the reflection matrices.
 Our calculation is simple such that one can extend it to systems containing inhomogeneous layers as good candidates for designing nanomachines. We find that the Casimir force is proportional to $L^{-5}$ where $L$ is the thickness of the inner layer.
\end{abstract}

\pacs{03.65.Nk, 42.50.Lc}

\maketitle
\section{Introduction}
 The Casimir effect~\cite{l1} resulting from modifying of the vacuum fluctuations due to the insertion of the boundaries has entered a new era of novel accurate measurements.
 According to the rapid progress of the nanotechnology and the introduction of the Casimir force offering new possibility for designing nanomechanical systems~\cite{l5,l5-2,l5-3,l5-4}, it will be useful to utilize multilayers in the investigating systems to evaluate Casimir energy.
 Having a repeating arrangement of thin layers of two different materials, this class of materials plays an important role in different branches of science and technology.
 Studying the properties of multilayers continues to grow exponentially due to their application at the biological interfaces for controlling drug molecules' release and permission as an instance~\cite{l5-1}.
 The non-pairwise addition effect in the van der Waals-Lifshitz interaction has been investigated in a multilayer system in~\cite{l6,l7}.
 They have explored the effect of the presence of other layers in the multilamellar geometries applying an algebra of $2\times2$ matrices resulting in the interesting consequence of the nonadditivity of the Casimir effect.
 Considering two periodic dielectric gratings, Lambrecht and Marachevsky have presented an exact calculation to obtain the Casimir energy~\cite{l9}. It is worth mentioning that comparing their theoretical calculation and the measurement performed by Chan et al.~\cite{l10} for grating-sphere geometry manifests a meaningful agreement.
 Evaluating a modal approach in~\cite{l11} -based on the scattering approach~\cite{l12,l12-1,l12-2,l12-3,l12-4}-
  the researchers have investigated the finite temperature Casimir interaction force between two periodic nanostructures and they have also illustrated the flexibility of this formalism.
 The Casimir energy between a plate and a nanostructured surface at arbitrary temperature has been calculated in the framework of the scattering theory in~\cite{l13} and as a significant consequence of this investigation it is illustrated that for grating geometries the contribution of the thermal part of the Casimir energy is intensified at small separation distances.
 Modeling grating as a dielectric function depending on the space and frequency as well as using variable phase method, Graham in~\cite{l14} has presented an appropriate approach to investigate the Casimir effect for gratings with deep corrugations.
 \par
\begin{figure}
\includegraphics[width=0.8\columnwidth]{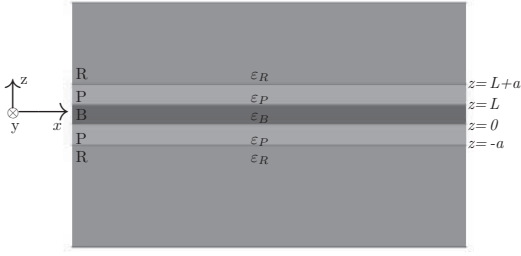}
\caption{A system composed of two half spaces of a material characterized with the permittivity $\varepsilon_{R}(\omega)$ as substrates within three alternating homogenous layers described by permittivities $\varepsilon_{B}(\omega)$ and $\varepsilon_{P}(\omega)$.}
\label{fig1}
\end{figure}

 In this work considering two half-space mediums as substrates and a symmetric array composed of three homogenous layers of alternating materials in between (see Fig.\ref{fig1}), we present reflected and transmitted amplitudes with the efficient approach of~\cite{l4-1}.
 With respect to the properties of a dielectric, we regard the permittivity of the materials to depend on the frequency in order to assume a real-world system.
 There are different approaches to obtain the Casimir energy, but not all efficient in calculating that for multilayers, gratings and other similar cases. The divergency occurring in calculating the Casimir energy dose not appear in the scattering formalism and besides with the advent of this approach, efficient evaluation of the Casimir energy for a wide variety of geometries, materials, and external conditions has been made possible~\cite{l15}. According to these features, we initiate from a scattering approach with functional determinant representation at finite temperature to determine the Casimir energy for the configuration of Fig.\ref{fig1}.
 In a qualitative sense, we obtain the value of the free energy that agrees with the amount evaluated by Podgornik and his colleagues in~\cite{l7} using a different approach.
 The ability of the scattering approach which allows one to consider non-trivial geometries
at finite temperature together with a realistic
description of the material properties, permits us to calculate the Casimir energy for a configuration with a periodic structure in its layers (i.e. inhomogeneous configuration). This approach makes it possible to investigate such a configuration in future with the purpose of utilizing that in designing nanomachines.

\section{Constructing reflection matrices}
Let us consider two half-space mediums with a symmetric array composed of three layers of alternating materials in between, as depicted in Fig.\ref{fig1}. $B$, i.e. the inner homogeneous region, is $0<z<L$ with the permittivity $\varepsilon_{B}(\omega)$. Over this, there exists another homogeneous region, $L<z<L+a$, labeled $P$ with the permittivity $\varepsilon_{P}(\omega)$ and above that we have an $R$-labeled homogeneous region characterized by the permittivity $\varepsilon_{R}(\omega)$. Below region $B$, symmetric to the upper half of this system, in the region $-a<z<0$ there is another medium $P$ described with the permittivity $\varepsilon_{P}(\omega)$ and under that we have another $R$-region characterized by $\varepsilon_{R}(\omega)$.

We consider a plane wave propagating along the $z$-axis reflected and transmitted through the nearby layers.
Considering that this configuration is $t$, $x$ and $y$ invariant, we can extract a factor $e^{i(k_{x}x+k_{y}y-\omega t)}$ from all of the components of the electromagnetic fields.
In the absence of the upper half of Fig.\ref{fig1}, we introduce the $y$-components of the electric and magnetic fields in the medium $B$ as:
\begin{equation} \label{eq1}
\left\{\begin{matrix}
E_{y}^{B}=I_{B}^{e}e^{-i\gamma_{B}z }+R_{B}^{e}e^{i\gamma_{B}z }\\
H_{y}^{B}=I_{B}^{h}e^{-i\gamma_{B}z }+R_{B}^{h}e^{i\gamma_{B}z }
\end{matrix}\right.
\end{equation}
where $I^{e/h}_{B}$ and $R^{e/h}_{B}$ are the incident and reflection coefficients of the electric (magnetic) waves. $\gamma_{B}$ indicates the $z$ component of the wave vector in the medium $B$, which can be introduced for the medium $j=B$, $P$ or $R$ as
\begin{equation}\label{eq2}
\gamma _{j}=i\bigg (-\varepsilon _{j}(\omega )\frac{\omega ^{2}}{c^{2}}+k_{x}^{2}+k_{y}^{2} \bigg )^{1/2}
\end{equation}
In this relation $\omega$ is the frequency, $c$ denotes the speed of light and $k_{x}$ and $k_{y}$ are the longitudinal components of the wave vectors. Regarding the Drude model, we assume the relative permittivity of the region $\varepsilon_{j}(\omega)$ as function of frequency including dissipation.
\par
Now we determine the reflection coefficients of Eq.~(\ref{eq1}). In this regard we introduce the $y$-component of the fields in the $P$ layer as
\begin{equation}\label{eq3}
\left\{\begin{matrix}
E_{y}^{P}=T_{P}^{e}e^{-i\gamma_{P}z }+R_{P}^{e}e^{i\gamma_{P}z }\\
H_{y}^{P}=T_{P}^{h}e^{-i\gamma_{P}z }+R_{P}^{h}e^{i\gamma_{P}z }
\end{matrix}\right.
\end{equation}
\\
including a reflection and also a transmission contribution for both of the fields. Considering that region $R$ is extremely large in comparison with other regions of the system, we can write the $y$-components of the electric and magnetic fields in this region as the following simple form
\begin{equation}
\left\{\begin{matrix} \label{eq4}
E_{y}^{R}=T_{R}^{e}e^{-i\gamma_{R}(z+a) }\\
H_{y}^{R}=T_{R}^{h}e^{-i\gamma_{R}(z+a) }
\end{matrix}\right.
\end{equation}
where $T^{e}_{R}$ and $T^{h}_{R}$ are the transmitted electric and magnetic coefficients through the medium $R$, respectively. Due to the thickness of the $P$ layer and for the sake of convenience of imposing the boundary condition, a translation of magnitude $a$ becomes obvious in this relation.
We treat the $y$-components of the electromagnetic fields of different mediums in detail. Now with the aid of the Maxwell's curl equations, the $x$-components of the fields in the medium $j$ are given by
\begin{equation} \label{eq5}
\left\{\begin{matrix}
E_{x}^{j}=\frac{ik_{y}}{\varepsilon_{j}(\omega )\omega ^{2}/c^{2}-k_{y}^{2}}\partial _{x} E_{y}^{j}-\frac{i\omega\mu _{0}}{\varepsilon_{j}(\omega ) \omega ^{2}/c^{2}-k_{y}^{2}}\partial _{z} H_{y}^{j}\\
H_{x}^{j}=\frac{ik_{y}}{\varepsilon_{j}(\omega )\omega ^{2}/c^{2}-k_{y}^{2}}\partial _{x} H_{y}^{j}+\frac{i\omega\varepsilon  _{j}(w)\varepsilon _{0}}{\varepsilon_{j}(\omega )\omega ^{2}/c^{2}-k_{y}^{2}}\partial _{z} E_{y}^{j}
\end{matrix}\right.
\end{equation}
where $\varepsilon_{0}$ and $\mu_{0}$ are the physical constants corresponding to the vacuum permeability and permittivity and $\vec{H}$ denotes the magnetic field strength.
\par
Now we want to obtain the reflection coefficients of medium $B$, while we have to consider the upper half of Fig.\ref{fig1} as well as the lower half. In this regard, we made the following simple change in the variables of Eqs.~(\ref{eq1}), (\ref{eq3}) and (\ref{eq4}) analogous to an $L$ translation along the positive direction of the $z$-axis and an incident onto the upper surface
\begin{equation} \label{eq6}
\begin{matrix}
z\rightarrow L-z
\end{matrix}
\end{equation}
and therefore, at the interface of medium $B$ and its upper $P$ medium, the $y$-components of the electric and magnetic fields can be written as

\begin{equation} \label{eq1-1}
\left\{\begin{matrix}
E_{y}^{B}=I_{B}^{e}e^{-i\gamma_{B}(L-z) }+R_{B}^{e}e^{i\gamma_{B}(L-z) }\\
H_{y}^{B}=I_{B}^{h}e^{-i\gamma_{B}(L-z) }+R_{B}^{h}e^{i\gamma_{B}(L-z) }
\end{matrix}\right.
\end{equation}
Here $I_{B}^{e}$ and $I_{B}^{h}$ are the incident coefficients of the electric and magnetic fields in the medium $B$, respectively.
$R_{B}^{e}$ and $R_{B}^{h}$ that are the reflection coefficients of the electric and magnetic fields play an important role in constructing the reflection matrix of the upper half of Fig.\ref{fig1}.
For medium $P$, one can write the $y$-components of the electric and magnetic fields analogous to Eq.~(\ref{eq3}) as the following
\begin{equation}\label{eq3-1}
\left\{\begin{matrix}
E_{y}^{P}=T_{P}^{e}e^{-i\gamma_{P}(L-z) }+R_{P}^{e}e^{i\gamma_{P}(L-z) }\\
H_{y}^{P}=T_{P}^{h}e^{-i\gamma_{P}(L-z) }+R_{P}^{h}e^{i\gamma_{P}(L-z) }
\end{matrix}\right.
\end{equation}
\\
where $T_{P}^{e/h}$ denotes transition from medium $P$ to medium $R$ in the upper half of Fig.\ref{fig1} and $R_{P}^{e/h}$ describes the reflection from the interface of regions $P$ and $R$.
Finally for the $y$-components of the electric and magnetic fields of medium $R$ we have
\begin{equation}
\left\{\begin{matrix} \label{eq4-1}
E_{y}^{R}=T_{R}^{e}e^{-i\gamma_{R}(L+a-z) }\\
H_{y}^{R}=T_{R}^{h}e^{-i\gamma_{R}(L+a-z) }
\end{matrix}\right.
\end{equation}
where $T_{R}^{e}$ and $T_{R}^{h}$ are the transition coefficients of medium $R$. The $x$-components of these fields are given by Eq.~(\ref{eq5}).

Imposing the continuity boundary condition on the introduced longitudinal components of both electric and magnetic fields on the interface of the layers at $z=0$, we have
\begin{widetext}
\begin{equation}
   \left\{ \begin{array}{ll}
      \frac{-k_x k_y}{\varepsilon_{B}(\omega )\omega ^{2}/c^{2}-k_{y}^{2}}(I_{B}^{e}+R_{B}^{e})-\frac{\omega\mu _{0} \gamma_{B}}{\varepsilon_{B}(\omega )\omega ^{2}/c^{2}-k_{y}^{2}} (I_{B}^{h}-R_{B}^{h})

      = \frac{-k_x k_y}{\varepsilon_{P}(\omega )\omega ^{2}/c^{2}-k_{y}^{2}}(T_{P}^{e}+R_{P}^{e})-\frac{\omega\mu _{0} \gamma_{P}}{\varepsilon_{P}(\omega )\omega ^{2}/c^{2}-k_{y}^{2}} (T_{P}^{h}-R_{P}^{h})
      \\
      \\
      I_{B}^{e}+R_{B}^{e}=T_{P}^{e}+R_{P}^{e}
      \\
      \\
      \frac{-k_x k_y}{\varepsilon_{B}(\omega )\omega ^{2}/c^{2}-k_{y}^{2}}(I_{B}^{h}+R_{B}^{h})+\frac{\omega \varepsilon_{B}\varepsilon_{0} \gamma_{B}}{\varepsilon_{B}(\omega )\omega ^{2}/c^{2}-k_{y}^{2}} (I_{B}^{e}-R_{B}^{e})

      = \frac{-k_x k_y}{\varepsilon_{P}(\omega )\omega ^{2}/c^{2}-k_{y}^{2}}(T_{P}^{h}+R_{P}^{h})+\frac{\omega \varepsilon_{P}\varepsilon_{0}\gamma_{P}} {\varepsilon_{P}(\omega )\omega ^{2}/c^{2}-k_{y}^{2}} (T_{P}^{e}-R_{P}^{e})
      \\
      \\
      I_{B}^{h}+R_{B}^{h}=T_{P}^{h}+R_{P}^{h}
    \end{array}
     \right. \label{eq6-1}
\end{equation}
\end{widetext}
We exert the mentioned boundary condition on the surface $z=-a$ to obtain the following set of equations
\begin{widetext}
\begin{equation}
   \left\{ \begin{array}{ll}
      \frac{-k_x k_y}{\varepsilon_{P}(\omega )\omega ^{2}/c^{2}-k_{y}^{2}}(T_{P}^{e}\, e^{i \gamma_{P}a } +R_{P}^{e}\, e^{-i \gamma_{P}a })-\frac{\omega\mu _{0} \gamma_{P}}{\varepsilon_{p}(\omega )\omega ^{2}/c^{2}-k_{y}^{2}} (T_{P}^{h}\, e^{i \gamma_{P}a }-R_{P}^{h}\, e^{-i \gamma_{P}a })

      = \frac{-k_x k_y}{\varepsilon_{R}(\omega )\omega ^{2}/c^{2}-k_{y}^{2}}T_{R}^{e}-\frac{\omega\mu _{0} \gamma_{R}}{\varepsilon_{R}(\omega )\omega ^{2}/c^{2}-k_{y}^{2}} T_{R}^{h}
      \\
      \\
      T_{P}^{e}\,e^{i \gamma_{P}a }+R_{P}^{e}\,e^{-i \gamma_{P}a }=T_{R}^{e}
      \\
      \\
      \frac{-k_x k_y}{\varepsilon_{P}(\omega )\omega ^{2}/c^{2}-k_{y}^{2}}(T_{P}^{h}\,e^{i \gamma_{P}a }+R_{P}^{h}\,e^{-i \gamma_{P}a })+\frac{\omega \varepsilon_{P}\varepsilon_{0} \gamma_{P}}{\varepsilon_{P}(\omega )\omega ^{2}/c^{2}-k_{y}^{2}} (T_{P}^{e}\,e^{i \gamma_{P}a }-R_{P}^{e}\,e^{i \gamma_{P}a })

      = \frac{-k_x k_y}{\varepsilon_{R}(\omega )\omega ^{2}/c^{2}-k_{y}^{2}}T_{R}^{h}+\frac{\omega \varepsilon_{R}\varepsilon_{0}\gamma_{R}} {\varepsilon_{R}(\omega )\omega ^{2}/c^{2}-k_{y}^{2}} T_{R}^{e}
      \\
      \\
      T_{P}^{h}\,e^{i \gamma_{P}a }+R_{P}^{h}e^{-i \gamma_{P}a }=T_{R}^{h}
    \end{array}
     \right. \label{eq6-2}
\end{equation}
\end{widetext}

Considering that the incident coefficients of the electric and magnetic fields are indeed inputs of our problem, we first assume $I_{B}^{e}=1$ and $I_{B}^{h}=0$ to solve this system of equations with the purpose of obtaining desired variables $R_{B}^{e}$ and $R_{B}^{h}$ as the elements of the first column of the reflection matrix of the inner medium analogous to the lower half of Fig.\ref{fig1}, i.e. $R_{l} (\omega)$. Then assuming $I_{B}^{e}=0$ and $I_{B}^{h}=1$, one is able to obtain the other column of this reflection matrix.
 We construct the reflection matrix of region $B$ for the electromagnetic wave reflected from the lower nearby layers with the following recipe
\begin{equation} \label{eqRu}
R_{l}(\omega)=\left(
                \begin{array}{cc}
                  R_{B}^{e}(I_{B}^{e}=1) & R_{B}^{e}(I_{B}^{h}=1) \\
                  R_{B}^{h}(I_{B}^{e}=1) & R_{B}^{h}(I_{B}^{h}=1) \\
                \end{array}
              \right)
\end{equation}
where, as an instance, $R_{B}^{e}(I_{B}^{e}=1)$ refers to assuming $I_{B}^{e}=1$ along with $I_{B}^{h}=0$ in Eqs.~(\ref{eq6-1}) and (\ref{eq6-2}) to obtain $R_{B}^{e}$ in these system of equations.
To construct $R_u (\omega)$ that is the reflection matrix due to the upper half of our system, we have performed the same procedure for the tangential components of the electromagnetic fields corresponding to the upper half of the system introduced in Eqs.~(\ref{eq1-1}),~(\ref{eq3-1}) and (\ref{eq4-1}).

\section{Calculating Casimir force in region $B$}
\begin{figure}
\includegraphics[width=0.8\columnwidth]{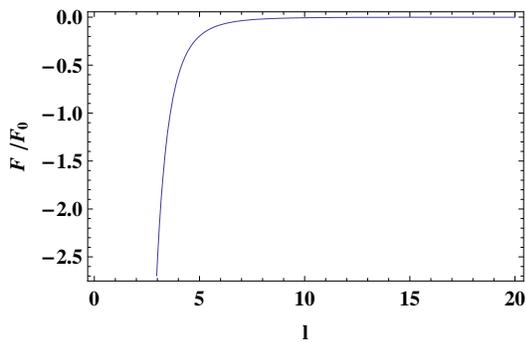}
\caption{Plot of the Casimir interaction force as a function of the dimensionless parameter $l=L/a$ in units of $F_0=10^{-11}\:N$ for the configuration of Fig.\ref{fig1} at room temperature. Here we have considered water and lipid for the alternating layers with $a=4\:nm$ as the thickness of medium $B$. As it is expected for large values of the dimensionless parameter $l$, this Casimir force converges rapidly due to the feature of the Casimir effect.}
\label{fig2}
\end{figure}

In this section we will use the scattering approach to obtain the Casimir force for the system composed of two substrates within three homogeneous alternating materials (see Fig.\ref{fig1}). For this purpose we construct a matrix $\textit{M}$ depicting a full-round trip through medium $B$ as
\begin{equation} \label{eq7}
\textit{M}(\omega )=R_{u}(\omega )e^{i\gamma_{B}L }R_{l}(\omega )e^{i\gamma_{B}L }
\end{equation}
where $L$ denotes the thickness of medium $B$, $e^{i \gamma_{B}L}$ refers to the free photon propagation in this medium and $R_{u}(\omega)$ and $R_{l}(\omega)$ are the upper and lower reflection matrices that we have presented in the previous section. Assuming $\omega=i \zeta_{n}$ with $\zeta_{n}=\frac{2\pi n k_{B} T}{\hbar}$, we utilize the Matsubara frequency to obtain the finite temperature Casimir force per unit area from the following equation
\begin{equation} \label{eq8}
\mathcal{F}(L)=2\pi k_{B} T {\sum_{n=0}^{\infty}} {'}\int_{-\infty }^{+\infty }\int_{-\infty }^{\infty }tr[(1-\textit{M}_{n} )^{-1}\partial _{L}\textit{M}_{n}]dk_{x}dk_{y}
\end{equation}
where $k_{B}$ is the Boltzmann constant, $T$ denotes the temperature and $n$ is an integer (see~\cite{l9,l13} and references there in). To proceed, we assume materials of regions $B$, $P$ and $R$ to be water, lipid and water in order to follow the configuration presented in~\cite{l7} closely. Performing the calculations for this special case as it is illustrated in Fig.\ref{fig2}, we obtain the Casimir force in the $SI$ system of units as
\begin{equation} \label{eq9}
F(l)= -6.176\times10^{-9}/\,l^{5}
\end{equation}
in which $l$ is a dimensionless parameter introduced as $l=L/a$ and $a$ is the thickness of medium $P$.
Consequently the Casimir energy of this configuration is proportional to $L^{-4}$ which is the same as the energy dependence obtained in~\cite{l7} for the investigated configuration. We should emphasize that in the asymptotic limit $L\gg a$, assuming $a=4\:nm$ and $L=100\:nm$ a free energy equal to $1.5\times10^{-23}\:j$ has been obtained in~\cite{l7} which is in a good agreement with the value of the Casimir energy that we have computed for the parameter $l=L/a=100/4$.

\section{Conclusion}
In this paper applying the scattering approach along with the Matsubara formalism, we have presented a calculation of the finite temperature Casimir force for a system composed of two half space materials with a three-layer array of alternating materials between them.
 We have considered Drude model for dielectric functions in order to take the materials'e dissipation and frequency dependence into account. Introducing longitudinal components of the electromagnetic fields in different regions of Fig.\ref{fig1}, we have imposed the continuity boundary condition on these fields to construct the reflection matrices. We have considered mediums of water and lipid to follow the configuration presented in~\cite{l7}, closely. Assuming $l=L/a=100/4$ with $L$ and $a$ as the thickness of mediums $B$ and $P$, we have computed a Casimir energy equal to $1.5\times10^{-23}\:j$ at room temperature. This result fits with the value of the energy corresponding to $a=4\:nm$ and $L=100\:nm$ in the $L\gg a$ limit of~\cite{l7}. This agreement illustrates the validity of our simple and transparent procedure which is fortunately able to evaluate the Casimir energy for structures with inhomogeneous layers which are recognized as good candidates for engineering Casimir effect with the purpose of designing miniaturized systems.
\par


\end{document}